\documentstyle[12pt,fleqn,epsfig]{article}

\begin{document}

%\documentstyle[12pt,epsfig]{article}
%\begin{document}
%\newcommand{\ba}{\begin{array}}
%\newcommand{\ea}{\end{array}}
%\title{\bf Nuclear binding energies (I): Global structure and local
%shell-model correlations}

\newcommand{\1}{\'{e}\'{e}n}
\newcommand{\be}{\begin{equation}}
\newcommand{\ee}{\end{equation}}
\newcommand{\ba}{\begin{array}}
\newcommand{\ea}{\end{array}}
\newcommand{\sub}[3]{ \mbox{ $ #1_{\mbox{\scriptsize #2}}^{#3} $ } }
\newcommand{\fig}[4]{ \vspace*{#1cm} \begin{center} {\footnotesize {\bf Fig. #2} #3 \cite{#4} .} \end{center} }
\newcommand{\Fig}[3]{ \vspace*{#1cm} \begin{center} {\footnotesize {\bf Fig. #2} #3  .} \end{center} }
\newcommand{\Table}[3]{ \vspace*{#1cm} \begin{center} {\footnotesize {\bf Table #2} #3  .} \end{center} }
\newcommand{\ster}{ \begin{center} * \ \ * \ \ * \end{center} }

\title{\bf Proton-neutron quadrupole interactions: an effective contribution to the
pairing field}

\vspace{2cm}

\author{\bf R.Fossion, C.De Coster\thanks{Postdoctoral fellow of the Fund for 
Scientific Research-Flanders (Belgium)}, J.E.Garcia-Ramos\thanks{Visiting postdoctoral fellow of the Fund for Scientific Research-
Flanders (Belgium)},\\
\bf and K.Heyde\thanks{Present address: EP-ISOLDE, CERN, CH-1211 Geneva, Switzerland}\\
\em Department of Subatomic and Radiation Physics,\\
\em Proeftuinstraat,86 B-9000 Gent, Belgium}

\date{ }
\maketitle

\begin{abstract}

We point out that the proton-neutron energy contribution, for low multipoles
(in particular for the quadrupole component), effectively renormalizes the
strength of the pairing interaction acting amongst identical nucleons filling
up a single-j or a set of degenerate many-j shells. We carry out the
calculation in lowest-order perturbation theory. We perform a study of this
correction in various mass regions. These results may have implications for
the use of pairing theory in medium-heavy nuclei and for the study of 
pairing energy corrections to the liquid drop model when studying nuclear
masses.

\end{abstract}

\newpage

\section{Introduction}

The pairing force, which expresses in the most succint way the preference
of nucleon pairs to become bound into $J^{\pi}=0^{+}$ states in the atomic
nucleus, has been widely used in many applications in the study of nuclear
structure properties \cite {Bruss77, Ring80, Talm93,Heyd94}. 
The special structure of the monopole pairing force has allowed to study
the classification of nucleons that occupy a set of single-particle
orbitals. The quasi-spin scheme \cite {Ring80,Talm93}, or, closely related 
the seniority quantum number $v$ \cite {Shal63}, leads to single-j and 
degenerate many-j
shell exactly solvable models.  They can be used as benchmarks to compare
with approximation methods. Moreover, use of monopole pairing forces allows to
determine an important energy contribution when evaluating total nuclear
binding energies.

Another important characteristic of the nucleon-nucleon effective interaction
acting inside atomic nuclei is expressed through the long-range components of
this interaction \cite{Bohr75}. The low multipoles and the quadrupole 
component, in
particular, are essential in generating low-lying nuclear collective phenomena.
They also contribute to the mean-field energy (binding energy in the
nuclear ground state, deformation properties,...) of the atomic nucleus.

These two components of the nucleon-nucleon effective force have formed 
a keystone to understand many facets of
nuclear structure: from few valence nucleons near to closed-shell configurations as
well as in those situations where many valence protons and neutrons are
actively present outside closed shells. They are essential ingredients of
any shell-model (or present large-scale shell-model) calculation, even though
in most of them either model interactions are constructed explicitely
or deduced from more realistic nucleon-nucleon potentials 
(see refs. \cite {Warb90,Pov94,Cau98,Ots98,Dean99} to cite just some recent
shell-model studies). 

It is our aim now, in the study of nuclear masses and two-neutron separation
energies $S_{2n}$,  to understand better the interplay
of global energy contributions (liquid-drop energy as 
determined from
simple models \cite {Wap58, Wap71} or from more sophisticated 
microscopic-macroscopic methods \cite{Moll81,Moll88, Moll95, Abou95, Lala99,
Gor01}) 
with local correlation effects. The latter contributions arise from 
specific nuclear structure effects taking pairing and low-multipole 
force components into account. Local correlation effects can come from
various origins such as (i) the presence of closed-shell discontinuities,
(ii) the appearance of local zones of nuclear deformation, and (iii) 
configuration mixing or shape mixing that will show up in the ground state
of the nucleus itself. A number of results have been published recently
on this topic \cite{Schwa01, Foss01}.

In the present paper, we study how monopole  pairing (that forms an
essential ingredient in all local energy correlations) can 
accomodate long-range forces and as such give rise to an effective pairing
force that can later be used when (i) applying pairing theory in order to 
study lowest-order
broken-pair excitations in medium-heavy nuclei, and (ii) see how, following
results obtained recently by Fossion et al. \cite {Foss01} concerning the 
study of two-neutron separation energies $S_{2n}$, the new pairing 
corrections on top of the liquid-drop energy could reproduce local binding
energy (and $S_{2n}$) variations even better.
   
In section 2, we
succintly indicate the results of monopole pairing in a single-j shell, and
we explicitely evaluate the proton-neutron quadrupole-quadrupole contribution
to the ground-state energy. Thereby we observe that its effect leads 
to an effective
pairing contribution. In section 3, we discuss applications in various
mass regions in order to estimate the effect of this renormalized 
'pairing-like'contribution. 

In the conclusion, we indicate that this extra
effect may well be interesting when studying nuclear structure 
properties not too far from closed shells in which the interactions 
amongst identical nucleons dominate the proton-neutron interaction effects.

\section{Shell-model correlations}
\label{sec-sm}

As discussed in the introductory section, the interplay of the monopole 
pairing force
and the proton-neutron low-multipole deformation-driving force components
are essential ingredients of any shell model calculation. The aim of the 
present paper is to point out that, to lowest order, the proton-neutron 
quadrupole part can be incorporated as a renormalization 
of the pairing force
strength. Of course, this means that applications will stick to regions near to
closed shells where the proton-neutron energy contribution is not the dominant
part. In open shells, with both valence protons and neutrons active,  one 
has to treat both components (monopole pairing and low-multipole
proton-neutron part) on equal footing. 

We first give a short reminder of 
the monopole pairing correlation energy part, considering the nucleons are
filling a single-j shell (or a set of degenerate-j shells). Secondly, 
we study the proton-neutron energy correction (quadrupole interaction) to be 
superposed to the monopole pairing part.

\subsection{Pairing energy corrections}
\label{sec-pairing}
\begin{figure}[hbt]
\caption{Spectrum for a pure pairing force within the $1h_{11/2}$
orbital, {\it i.e.}~for the $(1h_{11/2})^n$ spectrum, with seniority
$\upsilon=0$, $\upsilon=2$, $\upsilon=4$, and $\upsilon=6$ as a
function of the particle number $n$. The pairing strength $G$ has been
chosen as $G=0.25$ MeV \cite{Heyd94}.}
\begin{center}
\mbox{\epsfig{file=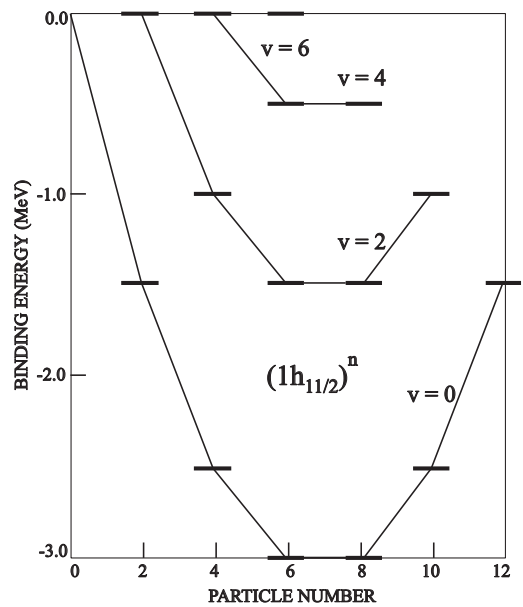,height=8cm,angle=0}}
\end{center}
\label{fig-pairing}
\end{figure}

The Hamiltonian, describing the most simple case of a monopole pairing force
between  $n$
identical valence nucleons interacting in a single-j shell, 
with a given strength $G$, is described as
\begin{equation}
\label{pair-ham}
\hat{H}=-G\sum_{m,m'>0}a^\dag_{jm} a^\dag_{j-m}
a_{j-m'} a_{jm'}(-1)^{2j+m+m'}. 
\end{equation} 
The ground state of such a system corresponds to a state where the
nucleons are coupled in pairs of angular momentum $J=0$.  As a
consequence the ground state has seniority $\upsilon=0$,
\begin{equation} 
\label{gs-pair}
|n,\upsilon=0\rangle=(S^+_j)^{n/2}|0\rangle,
\end{equation} 
where
\begin{equation}  
\label{s}
S^+_j= \frac{1}{\sqrt\Omega}\sum_{m>0}(-1)^{j+m} a^+_{jm} a^+ _{j-m}.
\end{equation} 
The binding energy of this system becomes
\begin{equation}
\label{pair-ener}
BE^{\scriptsize \mbox{pairing}} = \frac{G}{4} (2 \Omega - n + 2) n = 
G (\Omega - N + 1) N, 
\end{equation}
where $N$ is the number of valence nucleon pairs and $\Omega$ 
is the shell degeneracy, $\Omega=j+1/2$. In a system with valence 
protons and neutrons, interacting through monopole pairing forces between alike
nucleons, one has to consider the expression
(\ref{pair-ener}) for protons and for neutrons separately. 
As an illustration of this
interaction (see figure \ref{fig-pairing}), we shown the spectrum of a
monopole pairing force. From expression (\ref{pair-ener}),  the 
two-neutron separation energy can be easily deduced, with as a result
\begin{equation} 
\label{s2n-pair}
S_{2n}^{\scriptsize \mbox{pairing}} = G ( \Omega_\nu + 2 
- n_\nu) = G ( \Omega_\nu + 2 - 2 N_\nu),
\end{equation}
where $n_\nu$ is the number of valence neutrons, $N_\nu$ the number of 
valence neutrons pairs and $\Omega_\nu=j_\nu+1/2$ is the shell
degeneracy for neutrons.

%The main conclusion from this short section is that the pairing 
%contribution to the two-neutron separation energy $S_{2n}$ is linear in the number of neutrons.
%This simple result is in agreement with the specific linear stretches that
%are observed (see figures \ref{fig-s2n-vs} and \ref{fig-s2n-ca-xe-pb}) 
%for the various isotope series in medium-heavy and heavy nuclei,
%pointing toward the importance of the strong interactions
%amongst identical nucleons (neutrons in the present case) as main
%guide to understand the nuclear phenomena. 

\subsection{The quadrupole energy contribution}
\label{sec-quad}

%Besides the specific pairing correlations that determines an essential
%linear trend in the $S_{2n}$ values, we should study the effect of  
%low-multipole proton-neutron interaction on top of this linear
%dependence. 

As discussed before, we now evaluate the energy contribution in the 
ground state that is due to the proton-neutron 
quadrupole-quadrupole interaction.
The quadrupole proton-neutron Hamiltonian can be written as,
\begin{equation}
\label{ham-quad}
\hat H= \kappa \hat Q_\pi \cdot \hat Q_\nu, \label{quadrupole}
\end{equation}  
where $\hat Q$ is the quadrupole operator for protons and neutrons,
respectively, and $\cdot$ stands for the scalar product. The main
characteristic of the quadrupole operator is that it induces the breaking of
pairs, promoting pairs coupled to $J^\pi = 0^+$ into pairs coupled to 
$J^\pi = 2^+$, thereby changing the seniority quantum number
$\upsilon$ and causing a core-polarization effect
\cite{Hey87}. To a first approximation, the ground state of the system
will change from a condensate of proton and neutron pairs coupled to    
$J^\pi = 0^+$  into a superposition of the state of expression (\ref{gs-pair})
together with a new one where one-proton and one-neutron
pairs coupled to $J^\pi=2^+$ are induced,
\begin{equation} 
\label{gs-pair-q}
|N_\pi\otimes N_\nu\rangle=
|S_{j\pi}^{N_\pi}\otimes S_{j\nu}^{N_\nu};J=0\rangle + 
\xi|S_{j\pi}^{N_\pi-1}D_{j\pi}\otimes 
   S_{j\nu}^{N_\nu-1}D_{j\nu};J=0\rangle,  
\end{equation}  
where the operator
\begin{equation}        
\label{d}
D^+_j=\sum_{m>0}(-1)^{j-m}\langle j m j -m|2 0\rangle 
       a^\dag_{jm}a^\dag_{j-m},
\end{equation}  
will create a pair of nucleons coupled to $J=2$.

Taking into account that the quadrupole energy contribution is small with
respect to the monopole pairing interaction energy, the most 
straightforward way 
in order to fully determine the state vector in expression
(\ref{gs-pair-q}) is to use perturbation
theory. The mixing coefficient results as 
\begin{equation} 
\label{coef-mix}
\xi={\kappa\alpha\over\Delta E_\alpha},
\end{equation} 
where we use the shorthand notation 
\begin{equation}
\label{alpha}
\alpha= \langle S_{j\pi}^{N_\pi}\otimes S_{j\nu}^{N_\nu};J=0|   
         \hat Q_\pi \cdot \hat Q_\nu
        |S_{j\pi}^{N_\pi-1}D_{j\pi}\otimes 
         S_{j\nu}^{N_\nu-1}D_{j\nu};J=0\rangle ,
\end{equation} 
and define the energy difference
\begin{equation} 
\label{e-alpha}
\Delta E_\alpha=-E_{2^+_\pi} - E_{2^+_\nu},
\end{equation} 
in which $E_{2^+_\rho}$ ($\rho=\pi,\nu $) denotes the excitation energies 
of states
with one $J=2$ proton or neutron pair, respectively. 

The energy correction due to the quadrupole force is then given by 
evaluating the matrix element of the Hamiltonian (\ref{ham-quad}) using 
the state vector
(\ref{gs-pair-q}). For a system where the
forces are monopole pairing and quadrupole proton-neutron interactions
only, this results into the binding energy expression
\begin{equation} 
\label{be-pair-q1}
BE=BE^{\scriptsize \mbox{pairing}}+
     2{\kappa^2\alpha^2\over(-\Delta E_\alpha)},
\end{equation} 
where $BE^{\scriptsize \mbox{pairing}}$ corresponds to the result given in
expression (\ref{pair-ener}).

Due to the schematic structure of the states appearing in expression
(\ref{alpha}), it is now possible to obtain an explicit expression for
this matrix element \cite{Shal63}. 
In a first step, we decouple the proton and neutron parts
with the result
\begin{eqnarray} 
\label{alpha2}
\alpha&=&\frac{1}{\sqrt{5}}\langle(S_{j\pi})^{N_\pi};J=0\|
\hat Q_\pi \| (S_{j\pi})^{N_\pi-1}D_{j\pi};J=2\rangle\nonumber\\ 
&\times&\langle(S_{j\nu})^{N_\nu};J=0\|
\hat Q_\nu \| (S_{j\nu})^{N_\nu-1}D_{j\nu};J=2\rangle.
\end{eqnarray} 
Using the appropriate reduction formulae one arrives to the result
\begin{equation} 
\label{alpha3}
\alpha=\frac{1}{\sqrt{5}}\sqrt{N_\pi(\Omega_\pi-N_\pi)} 
\sqrt{N_\nu(\Omega_\nu-N_\nu)}\frac{\langle\hat Q_\pi\rangle 
\langle\hat Q_\nu \rangle}{\sqrt{(\Omega_\pi-1)(\Omega_\nu-1)}},
\end{equation} 
where,
\begin{equation} 
\label{q-pi}
\langle\hat Q_\pi\rangle=
\langle S_{j\pi}\|\hat Q_\pi \| D_{j\pi}\rangle=
\frac{2}{\sqrt{2 j_\pi+1}}\langle j_\pi\| \hat Q_\pi \| j_\pi\rangle ,
\end{equation} 
and
\begin{equation} 
\label{q-nu}
\langle\hat Q_\nu\rangle=
\langle S_{j\nu}\|\hat Q_\nu\|D_{j\nu}\rangle=
\frac{2}{\sqrt{2 j_\pi+1}}\langle j_\pi\| \hat Q_\nu \| j_\pi\rangle.
\end{equation} 
The expressions (\ref{q-pi}) and (\ref{q-nu}) can be even more
simplified if one uses harmonic oscillator wave functions
and uses the expression for the quadrupole operator, 
\begin{equation} 
\label{q-oper}
\hat Q =\sum_i r_i^2 Y_2(\theta_i,\varphi_i),
\end{equation} 
resulting into 
\begin{eqnarray} 
\label{q-osc}
\langle\hat Q_\rho\rangle&=&
\frac{2}{\sqrt{2j_\rho+1}}(N_{ho}+\frac{3}{2})
\langle j_\rho \| Y_{2}(\theta,\varphi)\|j_\rho \rangle \nonumber\\ 
&=&\sqrt{\frac{5}{\pi}}(N_{ho}+\frac{3}{2}) 
\frac{\frac{3}{4}-j_\rho(j_\rho+1)}
{\sqrt{(2j_\rho-1)j_\rho(j_\rho+1)(2j_\rho+3)}}.  
\end{eqnarray}  
Here $\rho=(\pi,\nu)$, $Y_{2}(\theta,\varphi)$ 
denotes the spherical harmonic with $L=2$
and $N_{ho}$ describes the number of quanta of the shell.

One finally  obtains a closed expression for the binding energy,
\begin{equation} 
\label{be-pair-q2}
BE=BE^{\scriptsize \mbox{pairing}}+
\frac{2}{5}\frac{\kappa^2}{(-\Delta E_\alpha)} 
\frac{N_\pi(\Omega_\pi-N_\pi)N_\nu(\Omega_\nu-N_\nu)}
{(\Omega_\nu-1)(\Omega_\pi-1)} 
\langle Q_\pi \rangle^2\langle Q_\nu\rangle^2.
\end{equation}

If we are interested in the study of binding energies within a set of isotopes
(thus $\Omega_{\pi}$ and $ N_{\pi}$ are fixed numbers), and by defining the
coefficient $\overline{C}$ as follows,
\begin{equation} 
\label{c}
\overline{C}=\frac{2}{5}\frac{\kappa^2}{(-\Delta E_\alpha)} 
\frac{N_\pi(\Omega_\pi-N_\pi)}{(\Omega_\nu-1)(\Omega_\pi-1)} 
\langle \hat Q_\pi \rangle^2\langle \hat Q_\nu \rangle^2,
\end{equation}
one obtains a ``correlated'' binding energy expression in the ground state 
\begin{equation}
\label{corren}
BE^{\scriptsize \mbox{correlated}}=BE^{\scriptsize \mbox{pairing}} + 
\overline{C}(\Omega_\nu-N_\nu +1)N_{\nu} - \overline{C}N_{\nu} \label{renormalised}.
\end{equation}
We can incorporate the proton-neutron quadrupole binding energy contribution as a renormalisation of the pairing strength in regions near closed shells where pairing dominates the proton-neutron quadrupole interaction. In these regions and for medium-heavy nuclei, which can be characterized by single-particle orbitals with large degeneracies $\Omega_\nu$ (see also discussion in sect.~3 and tables~1 and 3), the energy contribution for the third term in eq.~(\ref{renormalised}) is much smaller than the energy contribution for the second term at the beginning of the shell (small $N_\nu$) and up towards midshell ($N_\nu \approx \Omega_\nu / 2$). The total ``correlated'' binding energy then becomes to a good approximation,

\begin{equation}
\label{correno}
BE^{\scriptsize \mbox{correlated}} \approx ( G + \overline{C})(\Omega_\nu-N_\nu +1)N_{\nu} . \label{effective}
\end{equation}

So, one obtains a form, identical to the original monopole pairing energy 
expression, albeit with a new coupling strength. This equation can be applied to the region where detailed values for binding energies in the $Z \sim 40$ and $Z \sim 50$ nuclei (see table~3) are known.  

\vspace{0.5cm}

In the next section we shall evaluate, in some detail, the range of values
for the coefficients $G$ and  $\overline{C}$ in medium-heavy nuclei.
These results will show how good our idea is in reality, that proton-neutron quadrupole-quadrupole forces can be used to define an effective pairing force between identical nucleons.

An interesting result is that in evaluating the two-neutron 
separation energy, the quadratic part
drops out and one obtains a strictly linear behavior in $N_{\nu}$ 
with the result
\begin{equation} 
\label{s2n-quad}
S_{2n}^{\scriptsize\mbox{quadrupole}}(N_\nu)= 
( G+ \overline{C})(\Omega_\nu+2-2N_\nu) - \overline{C}.  
\end{equation}

This is essentially the same result as was obtained using a pure monopole
pairing
force (see expression (\ref{s2n-pair})) except for the small correction 
factor $\overline{C}$  (which we still have to prove). 

\vspace{0.5cm}

The former discussion which uses first order perturbation theory to determine the wave function (eq.~(\ref{gs-pair-q})) can be repeated for pure hexadecupole or higher multipole interaction Hamiltonians (using eq.~(\ref{quadrupole}), replacing $\hat{Q}_\rho$ by the appropriate multipole operator).  Each higher multipole separately will give a smaller energy contribution to the binding energy, so that only a limited number of multipoles will be important. 
Using higher order perturbation theory to determine the modified wave functions, extra energy contributions become are possible that come from different multipoles acting together. These higher order contributions result in a different $N_\nu$ dependence than the contributions from separately treated multipoles. These higher order effects will be of minor importance. In the present paper, we do not aim at carrying out a detailed shell model study - in that case the multipole Hamiltonian should rather be diagonalised - but we find it quite surprising that the lowest-order effect induced by
proton-neutron residual forces comes about as a 'renormalization' of 
the original monopole pairing force. We find this an interesting
observation. 

\section{The pairing and quadrupole strength: 
some specific applications and how well works the above approximation}
\label{sec-applic}

In the present section, we make a detailed study of the correction 
factor $\overline{C}$ , mainly concentrating 
on medium-heavy nuclei. We compare this effective pairing force with
the monopole pairing strength that is derived from standard parametrizations,
in this mass region.

\begin{figure}
\caption{Plot of the values for $\left<Q_\rho \right>$ and 
$\left< j_\rho \left\| Q_\rho \right\| j_\rho \right>$ for $j_\rho$ 
going from $3/2$ to $31/2.$ As one can see, $\left<Q_\rho \right>$ is 
nearly a constant for changing $j_\rho$.}
\begin{center}
\mbox{\epsfig{file=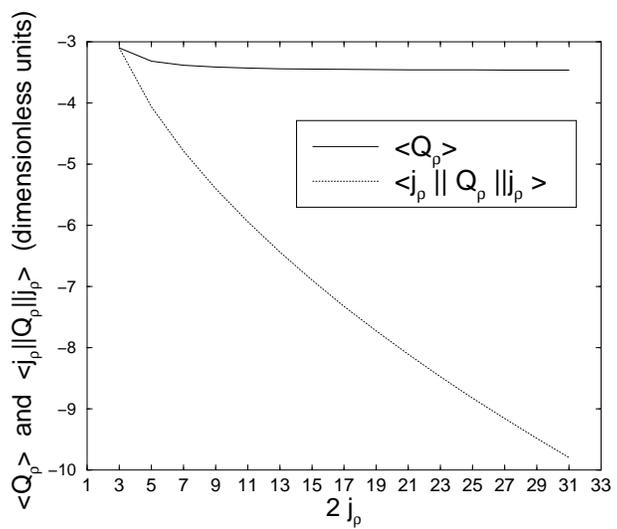,height=8cm,angle=-90}}
\end{center}
\label{fig-q}
\end{figure}

\begin{itemize}
\item{The monopole pairing strength $G$.}

An average value of $G$ for medium-mass and heavy nuclei is  
$25/A$ MeV \cite{Rowe70}. However, this value can significantly
change in different mass regions, ranging from $G=19/A$ MeV, for the
regions $Z\sim 40-50$, to $G=30/A$ MeV for Pb nuclei \cite{Kiss60}.

As an example, for $_{52}$Te isotopes ($A\approx 106-152$) the pairing
strength is $G\approx 0.15$ MeV, and for $_{42}$Mo isotopes 
($A\approx 84-128$) $G\approx 0.18$ MeV.

\item{The effective pairing strength $\overline{C}$, obtained from
the proton-neutron quadrupole force.}

In order to estimate the value of $\overline{C}$, using the method
discussed in section 2, we have to reduce the set of more realistic
single-particle orbitals that are not degenerate and are typical for 
a given mass region, into a large, degenerate single-j shell. 
In table \ref{tab-degen}, 
we indicate those values of $j$ as well as the degeneracies for different major
shells or subshells in the regions $Z\sim40$ and $Z\sim50$. As example
of isotopes with $Z\sim40$ we consider the $_{48}$Cd (N$_\pi=1$), 
$_{42}$Mo (N$_\pi=1$), $_{44}$Ru (N$_\pi=2$), and $_{46}$Pd
(N$_\pi=2$) nuclei . For the region with $Z\sim50$, we consider the 
$_{52}$Te (N$_\pi=1$), $_{54}$Xe
(N$_\pi=2$), $_{56}$Ba (N$_\pi=3$), and $_{58}$Ce (N$_\pi=4$) nuclei. 
For both shells we take as number of quanta N$_{ho}=4$, which covers a
shell from $40$ to $70$ nucleons and is in agreement with table
\ref{tab-degen}.

Another ingredient necessary in order to determine $\overline{C}$ (see
expression (\ref{c})) is the value of 
$\Delta E_\alpha=-E_{2^+_\pi} - E_{2^+_\nu}$
which can be estimated from experimental energy systematics of the 
first $2^{+}$ excitation energy \cite{Heyd94}. This results for the
$Z\sim50$ region into a maximum value of $\Delta E_\alpha \simeq - 2$ MeV 
\cite{Heyd86}
and for the $Z \sim 40$ region, a slightly higher value $\Delta
E_\alpha \simeq - 2.5$ MeV is obtained \cite{Hey87}.   

The final element in order to calculate $\overline{C}$ is the strength of
the proton-neutron interaction $\kappa$. There exist realistic values 
of the quadrupole
strength in the context of the IBM \cite{Iac87,Fra94}, $\kappa_0$,
that can be easily related with the value of $\kappa$ appearing in
expressions (\ref{ham-quad}) and (\ref{c}) through the following
relation \cite{Hey87}: 
\begin{equation}
\label{k-ibm-sm} 
\kappa=\frac{5\kappa_0\sqrt{\Omega_\pi\Omega_\nu}}
{\sqrt{\frac{2}{\Omega_\pi-1}}\sqrt{ \frac{2}{\Omega_\nu-1}} 
\langle j_\pi\|\hat Q_\pi\| j_\pi\rangle 
\langle j_\nu\|\hat Q_\nu\| j_\pi\rangle}.
\end{equation}

As can be verified from table \ref{tab-q} and noted in figure 
\ref{fig-q}, the reduced matrix elements
$|\langle j_\rho \| Q_\rho \| j_\rho \rangle|$ increase linearly with
$j_\rho$, but this variation is almost completely compensated through
the presence in the numerator of the factors $\Omega_\rho$.
This then results in values for $\kappa$ that are almost independent
of the value of $j_\rho$ ($\Omega_\rho$). On the other hand, the values
of $\langle \hat Q_\rho \rangle$ and $\langle\hat Q_\rho \rangle^2$ 
(the third factor in expression (\ref{c})) do not change dramatically 
with $j_\rho$ 
in the range $j_\rho= \frac32-\frac{31}{2}$. This is an important
outcome which implies that the value of $\overline{C}$ is not 
strongly dependent on the particular $j$ ($\Omega$) value of the
proton and/or neutron orbitals (degeneracies) that the nucleons are
occupying (state independence). This results into robust values
of $\overline{C}$.  

Having discussed the various elements that are necessary in order to 
derive a schematic, albeit rather realistic, estimate of $\overline{C}$,
we present, in
table \ref{tab-c}, the values for different nuclei in the
regions $Z\sim 40$ and $Z\sim 50$. We point out that two alternatives shell
closures have been used.  

Inspecting the results, as given in table \ref{tab-c}, it becomes clear
that the effective pairing correction,
stemming from the proton-neutron interaction, 
can be at maximal of the order of 10-15\% of the
regular monopole pairing strength. The precise values  
depend somewhat on the degeneracies of the proton and neutron
shell-model spaces that are used in order to describe the various series of
isotopes.
\end{itemize}

As a conclusion, on notices that
the quadrupole constant $\overline{C}$ is smaller than the monopole pairing 
constant $G$ by a factor varying in between 5 to 25. 
Therefore, quadrupole and low-multipole force components
give rise to contributions to the binding energy 
that exhibit the same structure (except for some smaller corrections)
as those resulting from monopole pairing forces solely. The particular
number dependence of the binding energy (see expression (\ref{correno})) 
occurs through the common factor $(\Omega - N +1) N $  with $ \Omega $ 
the degenarcy of the shell that is filling up with identical nucleons.

\section{Conclusion}

In the present paper, we have shown that pure monopole pairing can
accomodate long-range forces (we have studied in particular the case
of proton-neutron quadrupole forces but the extension to other low
multipoles is now straightforward) using perturbation theory, and as 
such give rise to an effective pairing force.  
This latter effect renormalizes the monopole pairing energy 
with an amount that can vary between 5 to 25\%.

This result came partly
as a surprise and did come up when we were studying the variation of nuclear
binding energies and, more in particular, two-neutron separation
energies $S_{2n}$ over a large region of nuclei (rare-earth mass
region, nuclei in the neutron-deficient Pb region \cite {Schwa01, Foss01}).  
In the above papers, we have studied the local correlation energy 
within the framework of the Interacting Boson Model (IBM)
\cite {Iac87,Fra94,Cas88} and outlined a prescription in order to 
derive nuclear masses within a single framework.

Recently, we have started the study of nuclear binding energies and two-neutron
separation energies $S_{2n}$ in medium-heavy nuclei (mass A=100-130
region) using the same concepts but also trying to evaluate the local
correlation energy taking into account the shell-model structure and
the residual interactions (pairing, quadrupole proton-neutron interactions)
\cite {Jose01}. In order to study these local energy corrections that
appear on top of the global energy (liquid drop energy, macroscopic-microscopic energy studies - see refs. given in the introduction), we
have studied in the present paper the modifications that proton-neutron
forces induce on the strict monopole pairing energy. We have shown that
the dependence on nucleon number (e.g. neutron number when studying
series of isotopes) for the effective pairing contribution is identical
to the nucleon number dependence of the monopole pairing force. In
going through a series of isotopes, changing the number of  neutron pairs,
$N_{\nu}$, the effective pairing force 
will likewise exhibit a neutron number dependence through the presence
of the factor $\overline{C}$ (see expression \ref{c}). 

Therefore, we aim at (i) applying pairing theory to study lowest-order
broken-pair excitations in medium-heavy nuclei, and (ii) see how, following
up on results obtained recently by Fossion et al. \cite {Foss01} 
concerning the 
study of two-neutron separation energies $S_{2n}$, the effective pairing 
corrections on top of the liquid-drop energy could reproduce local binding
energy (and $S_{2n}$) variations even better.

\section{Acknowledgements}

The authors are grateful to discussions with G.~Bollen, S.~Schwarz 
and A.~Kohl on very precise mass measurements and their importance for 
testing nuclear structure studies  that have lead to the present 
investigations. They are grateful to the referee for a number of
constructive remarks in order to improve on both form and content.
They thank the ``FWO-Vlaanderen'', NATO for a research grant CRG96-098  
and the IWT for financial support during this work. One of the authors 
(KH) is most grateful for the hospitality at ISOLDE/CERN and the support 
in finishing the present investigation.

\newpage
\noindent
{\bf Table Captions}\\

\begin{table}[hbt]
\caption{Degeneracies and associated single-j shell configurations
for regions $Z\sim40$ and $Z\sim50$.}
\begin{center}
\begin{tabular}{|l|l|l|}
%\hline
\hline
\multicolumn{3}{|c|}{$Z\sim40$}             \\
\hline
Shell      &Degeneracy       &$j$          \\
\hline
$Z=40-50$  &$\Omega_\pi=5$   &$j_\pi=9/2$ \\ 
$N=50-64$  &$\Omega_\nu=7$   &$j_\nu=13/2$ \\
$N=50-82$  &$\Omega_\nu=16$  &$j_\nu=31/2$ \\
\hline
\hline
\multicolumn{3}{|c|}{$Z\sim50$}             \\
\hline
Shell      &Degeneracy       &$j$          \\
\hline
$Z=50-82$  &$\Omega_\pi=16$  &$j_\pi=31/2$ \\ 
$Z=50-64$  &$\Omega_\pi=7$   &$j_\pi=13/2$ \\
$N=50-82$  &$\Omega_\nu=16$  &$j_\nu=31/2$ \\
%\hline
\hline
\end{tabular}
\end{center}
\label{tab-degen}
\end{table}

\clearpage
\begin{table}[hbt]
\caption{Values for $\left<Q_\rho \right>, \left<Q_\rho \right>^2$ 
and $\left< j_\rho \left\| Q_\rho \right\| j_\rho \right>$ 
(see expressions \ref{q-pi}, \ref{q-nu} and \ref{q-osc}) for
$j_\rho$ going from $3/2$ to $31/2.$}
\label{tab-q}
\begin{center}
\begin{tabular}{|c|c|c|c|} \hline
$j_\rho$ & $\left< Q_\rho \right>$ & $\left< Q_\rho \right>^2$ & $\left< j_\rho \| Q_\rho \| j_\rho \right>$  \\ \hline
3/2 &-3.10	&9.62	&-3.10 \\
5/2 &-3.31	&11.00	&-4.06 \\
7/2 &-3.38	&11.46	&-4.78 \\
9/2 &-3.41	&11.67	&-5.40 \\
11/2 &-3.43	&11.78	&-5.94 \\
13/2 &-3.44	&11.85	&-6.44 \\
15/2 &-3.44	&11.89	&-6.89 \\
17/2 &-3.45	&11.92	&-7.32 \\
19/2 &-3.45	&11.94	&-7.72 \\
21/2 &-3.45	&11.96	&-8.11 \\
23/2 &-3.46	&11.97	&-8.47 \\
25/2 &-3.46	&11.98	&-8.82 \\ 
27/2 &-3.46	&11.99  &-9.48 \\
31/2 &-3.46	&12.00	&-9.79 \\ \hline
\end{tabular}
\end{center}
\end{table}

\clearpage
\begin{table}[hbt]
\caption{Values of $\overline{C}$ ( in MeV) for different nuclei
 and different mass regions.}
\begin{center}
\begin{tabular}{|l|l|l|l|l|}
\hline
\multicolumn{5}{|c|}{$Z\sim40$}                            \\
\hline
Shells                         &Mo     &Cd    & Ru   & Pd   \\
\hline
$\Omega_\pi=5$, $\Omega_\nu=16$&0.0062 &0.0062&0.0093&0.0093\\
$\Omega_\pi=5$, $\Omega_\nu=7$ &0.0059 &0.0059&0.0079&0.0079\\
\hline
\hline
\multicolumn{5}{|c|}{$Z\sim50$}                            \\
\hline
Shells                                    &Te     &Xe    & Ba   & Ce  \\
\hline
$\Omega_\pi=16$, $\Omega_\nu=16$          &0.0091 &0.017 &0.0240&0.029\\
$\Omega_\pi=7\phantom{1}$, $\Omega_\nu=16$&0.0036 &0.006 &0.0073&0.0073\\
\hline
\end{tabular}
\end{center}
\label{tab-c}
\end{table}

\clearpage
%\newpage
\noindent
{\bf Figure Captions}\\

%\begin{figure}[hbt]
%\caption{Comparison between the pairing and quadrupole contributions
%to the binding energy in the case of $_{52}Te$. Observe the typical parabolic behaviour. 
%The quadrupole contribution is the most important near neutron midshell.}
%\begin{center}
%\mbox{\epsfig{file=plot_BE_Te.eps,height=8cm,angle=-90}}
%\end{center}
%\label{fig-pair-quad}
%\end{figure}

%\begin{figure}[hbt]
%\caption{$Z=40$ region. Comparison between the pairing and quadrupole 
%contributions to the two-neutron separation energy for $_{48}Cd$.}
%\begin{center}
%\mbox{\epsfig{file=plot_S2n_Cd.eps,height=8cm,angle=-90}}
%\end{center}
%\label{fig-pair-quad-40}
%\end{figure}

%\begin{figure}[hbt]
%\caption{$Z=50$ region. Comparison between the pairing and quadrupole 
%contributions to the two-neutron separation energy for $_{52}Xe$.}
%\begin{center}
%\mbox{\epsfig{file=plot_S2n_Xe.eps,height=8cm,angle=-90}}
%\end{center}
%\label{fig-pair-quad-50}
%\end{figure}

%\begin{figure}[hbt]
%\caption{Pairing and quadrupole $S_{2n}$ corrections added to the 
%global LDM two-neutron separation energy line of 
%equation --. }
%\label{fig-ldm-corr}
%\end{figure}

\end{document}